\documentclass[12pt]{article}
\usepackage{url}
\usepackage{amsfonts}
\usepackage{amsmath,amssymb,color}
\usepackage{graphicx}
\usepackage{enumitem}
\usepackage{multirow}
\usepackage{bm}
\usepackage[natbibapa]{apacite}
\usepackage{verbatim}
\usepackage{float}
\usepackage[toc,page]{appendix}
\usepackage[onehalfspacing]{setspace}
\usepackage{setspace}
\usepackage{adjustbox}
\usepackage{rotating}
\usepackage{subcaption}
\usepackage[]{quoting}
\newcommand{\ttt}{\boldsymbol \theta}

\newcommand{\argmax}{\operatornamewithlimits{arg\,max}}

\title{A Continuous-Time Dynamic Choice Measurement Model for Problem-Solving Process Data}

\author{Yunxiao Chen\\
London School of Economics and Political Science}
\date{}

\begin{document}
\maketitle

\doublespacing

\begin{abstract}
Problem solving has been recognized as a central skill that today's students need to thrive and shape their world. As a result, the measurement of problem-solving competency has received much attention in education in recent years.
A popular tool for the measurement of problem solving is simulated interactive tasks, which require students
to uncover some of the information needed to solve the problem through interactions with a computer-simulated environment.
A computer log file records a student's problem-solving process in details, including his/her actions and the time stamps of these actions.
It thus provides rich information for the measurement of students' problem-solving competency. 
On the other hand, extracting useful information from log files is a challenging task, due to its complex data structure.
In this paper, we show how log file process data can be viewed as a marked point process, based on which we
propose a continuous-time dynamic choice model.
The proposed model can serve as a measurement model for scaling students along the latent traits of problem-solving competency and action speed, based on data from one or multiple tasks.
A real data example is given based on data from Program for International Student Assessment 2012.
\end{abstract}	

\section{Introduction}

Problem-solving competency has been recognized as a central
skill that today's students need to thrive and shape their world \citep{griffin2014assessment,OECD2018future}.
As a result, the measurement of problem-solving competency has received much attention in education in recent years \citep[e.g.][]{OECD2012pisa,oecd2012literacy,OECD2016pisa,IEA2017timss,national2013technology}.
Computer-based simulated interactive tasks have become a popular tool for the measurement of problem-solving competency. They have been used in many national and international large-scale assessments, including the Program for International Student Assessment (PISA),
the International Assessment of Adult Competencies (PIAAC), and the National Assessment of Educational Progress (NAEP).
Comparing with static problems,  interactive tasks better reflect the nature of problem solving in real life by
requiring students to uncover some of the information needed to solve the problem through interactions with a computer-simulated environment, while static problems disclose all information at the outset.

For simulated tasks, data are available not only for the final outcome of problem solving (success/failure), but also the entire problem-solving process recorded by computer log files. A computer log file contains events during a student's problem-solving process
(i.e., actions taken by the student) and the time stamps of these events, where the final outcome is completely determined by the problem-solving process. Therefore, problem-solving process data should contain
more information about one's problem-solving competency than the final outcome.
However, due to the complex structure of log file process data,
it is unclear how meaningful information can be extracted. Comparing with traditional multivariate data that are commonly encountered in social and behavioral sciences, such as testing data and survey data, computer log file data are highly unstructured. Different students can have completely different computer log files, with different events occurring at different time points.  

In this paper, we propose a probabilistic measurement model, called the Continuous-Time Dynamic Choice (CTDC) model, for extracting meaningful information from log file process data. We first provide a review of marked point process \citep{cox1980point}, a stochastic process whose realization takes the same form as log file process data.
 We then propose a parametrization of the marked point process, in which the occurrence of a future action and its time stamp depend on (1) the entire event history of problem solving, (2) person-specific characteristics, including the latent traits of problem-solving competency and action speed, and (3) task structure. In particular, we assume the choice of the next action is driven by a competency trait, while the time of action depends on a speed trait.
This model can be applied to data from 
one or multiple tasks.

The analysis of problem-solving process data has received much attention in recent years. 
A standard strategy to analyze such data
is based on summary statistics
defined by expert knowledge. These summary statistics are used for group comparison (e.g., comparing the success and failure groups) and/or multivariate analysis (e.g., factor analysis). Research taking this approach includes \cite{greiff2015computer}, \cite{scherer2015exploring}, \cite{greiff2016understanding}, and \cite{kroehne2018conceptualize}, among others.  Another type of analysis focuses on extracting
important features/latent features from process data. Along this direction, \cite{he2015identifying,he2016analyzing} took an n-gram approach to extract sequential features in data and screen out the important ones based on their predictive power of the problem-solving outcome.
\cite{xu2018latent} proposed a latent class model for finding latent groups among students based on log file data.
\cite{tang2019latent} proposed a multidimensional scaling approach to extracting latent features and show empirically that the extracted latent features tend to contain more information than the binary problem-solving outcome, in terms of out-of-sample prediction of related variables.
Besides these directions, \cite{chen2019statistical} proposed an event history analysis approach from a prediction perspective, studying how problem-solving process data can be used to predict the problem-solving outcome and duration.
However, all these approaches do not provide
a probabilistic measurement model that directly links together interpretable person-specific latent traits, the structure of problem-solving task, and log file process data. 

The proposed CTDC model is closely related to the Markov decision process (MDP) measurement model proposed by \cite{lamar2018markov} that is also used to measure student competency based on within-task actions. In particular, both the CTDC model and the MDP measurement model assume a dynamic choice model to characterize how the next action depends on the current status of the student (as a result of previous actions) and a person-specific competency latent trait.
In both models, a person with a larger latent trait level is more likely to choose a better action. 
However, there are several major differences between the two models. First, the MDP measurement model is only for the action sequences, without taking into account the time information of the actions that may also be informative.  On the other hand, by modeling log file data as a marked point process, the proposed framework is able to
make use of information from both the actions and their time stamps.
Second, the two models quantify the effectiveness of an action differently. The MDP measurement model follows a Markov decision theory framework.
It measures the effectiveness of an action given the student's current state by the value of a
Q-function (i.e., state-action value function) which is obtained by solving
an MDP optimization problem \citep[see][for the details of Markov decision process]{puterman2014markov}. This approach is possibly more useful for  complex tasks where the value of actions is hard to evaluate.
On the other hand, we focus on tasks for which
there exists a direct measure of action effectiveness based on their design. In fact,
for relatively simple tasks, such as those in large-scale assessments, it is often clear whether or not an action should be taken at each stage, which provides a measure of action effectiveness.
In particular, we demonstrate how a reasonable measure of action effectiveness can be constructed using a motivating example from PISA 2012, in which case the proposed approach is much easier to use.
Finally, the proposed model is developed under a general structural equation modeling framework that can
simultaneously analyze multiple tasks, while the MDP measurement model focuses on data from a single task.


The rest of the paper is organized as follows. In Section~\ref{sec:MPP}, we start with a motivating example from PISA 2012 and then provide a marked point process view of log file data. In Section~\ref{sec:model}, we propose a continuous-time dynamic choice (CTDC) measurement model under the marked point process framework, and discuss the estimation of model parameters.
In Section~\ref{sec:real}, the proposed model is applied to real data from PISA 2012, followed by a simulation study in Section~\ref{sec:sim}.
We end with discussions in Section~\ref{sec:diss}.


\section{Log File Data as a Marked Point Process} \label{sec:MPP}

\subsection{A Motivating Example}\label{subsec:exmp}

\begin{figure}
  \centering
  \includegraphics[scale = 1]{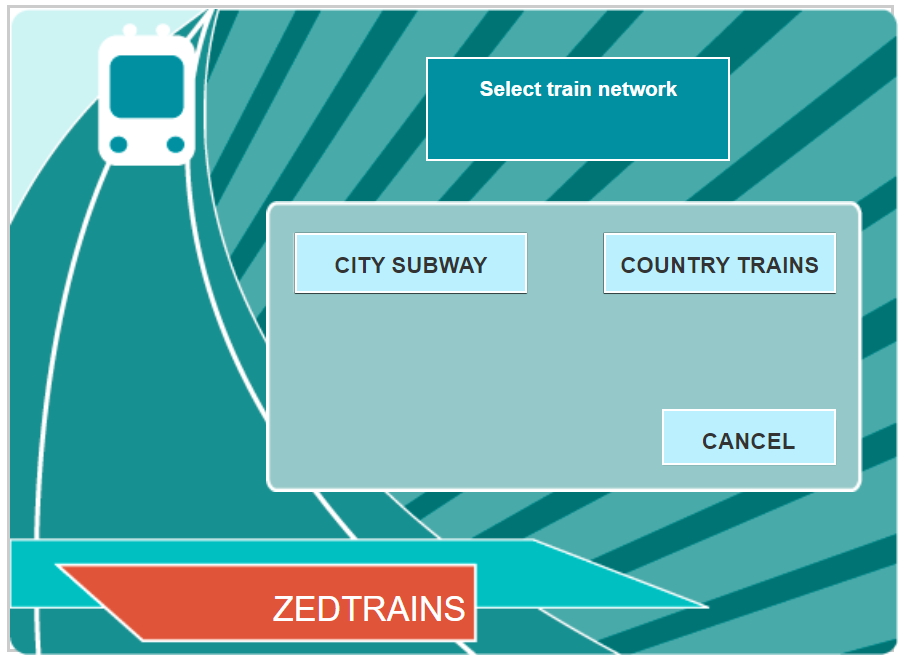}
  \caption{Screen shot of the starting screen of a problem-solving task
  from PISA 2012  about using a simulated automated ticketing machine.}\label{fig:ticket}
\end{figure}

To introduce the structure of log file process data, we start with a motivating example, which is
the second task  from a released
unit of PISA 2012 that contains three tasks.\footnote{The task is available online from \url{http://www.oecd.org/pisa/test-2012/testquestions/question5/}. }
This released unit is called TICKETS.
In this task, students
were asked to use a simulated automated ticketing machine to buy train tickets under certain constraints on the type of tickets. Figure~\ref{fig:ticket} provides a screen shot of the user interface for this unit of tasks. The instruction of the ticketing machine is given below.

\begin{quote}
\textit{``A train station has an automated ticketing machine. You use the touch screen on the right to buy a ticket. You must make three choices.
\begin{itemize}
  \item Choose the train network you want (subway or country).
  \item Choose the type of fare (full or concession).
  \item Choose a daily ticket or a ticket for a specified number of trips. Daily tickets give you unlimited travel on the day of purchase. If you buy a ticket with a specified number of trips, you can use the trips on different days.
\end{itemize}
The BUY button appears when you have made these three choices. There is a CANCEL button that can be used at any time BEFORE you press the BUY button."}
\end{quote}

In this task, the students were asked to find and buy the cheapest ticket that allows them to take four
trips around the city on the subway, within a single day. As students, they can use concession fares.
The accomplishment of the task requires multiple interactions between the student and the task interface. In particular,
the student needs to know the concession fare of a daily subway ticket and the concession fare of four individual subway tickets, by visiting the corresponding screens. Then the student needs to verify which of these is the cheapest ticket and make the purchase. We say the task is successfully solved if a student purchases four individual subway tickets in concession fare after
comparing its price to that of a daily subway ticket in concession fare.


This task is designed under the finite-state automata framework \citep{buchner1993finite,funke2001dynamic}, one of the most commonly used design for problem-solving tasks. In fact, it is one of the two design frameworks for all problem-solving tasks in PISA 2012. Tasks following the
finite-state automata design share a similar structure and  the proposed CTDC model can be applied to all such tasks.

\begin{figure}
  \centering
  \includegraphics[scale = 0.8]{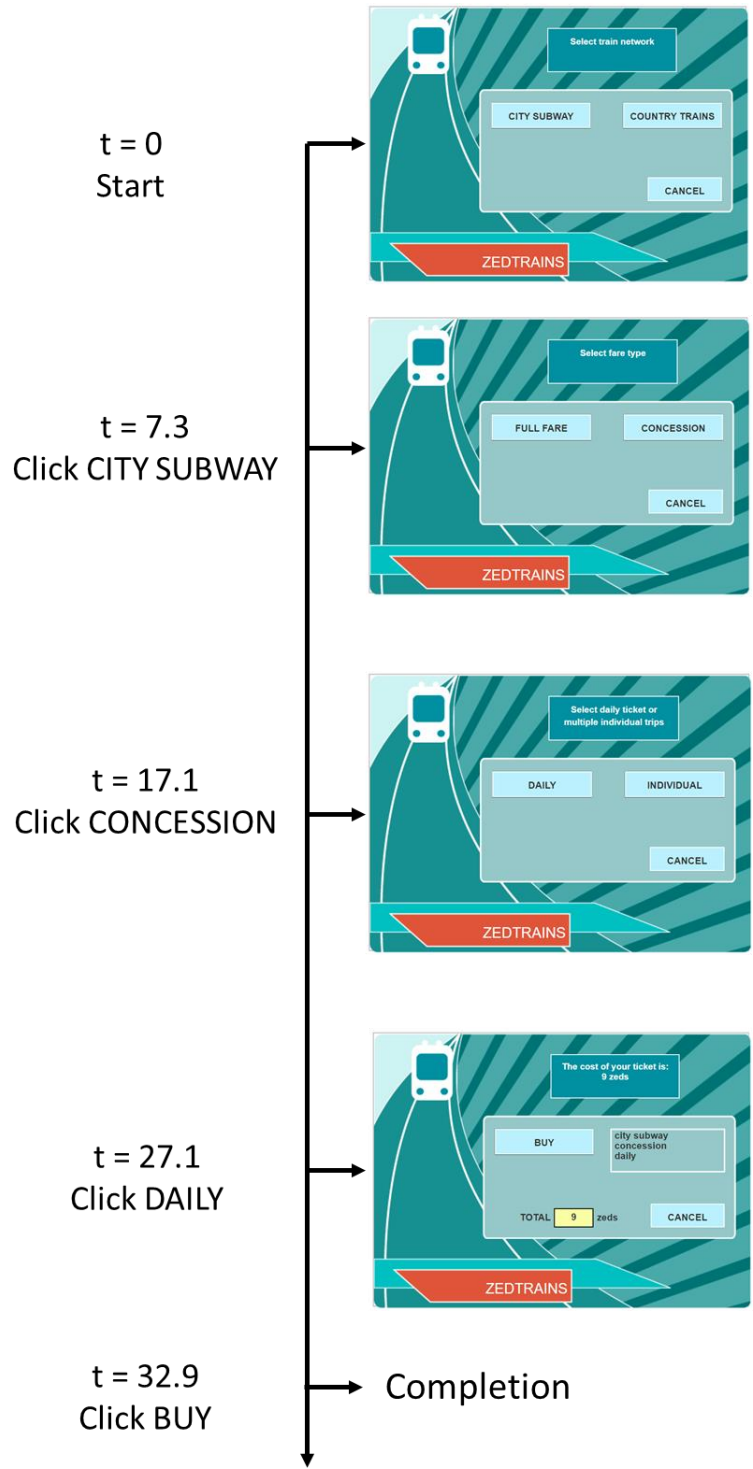}
  \caption{Visualization of a student's problem-solving process, where the starting time of the task is standardized to zero. }\label{fig:ticket2}
\end{figure}

The log file of a student solving a task is recorded using a long data format, with each row describing  an action and its time stamp. For an automata task,
a student's action can be represented by the resulting new state of the system. Figure~\ref{fig:ticket2} visualizes the problem-solving process of a student in PISA 2012 and Table~\ref{tab:ticket} shows the corresponding log file record.\footnote{The raw data are available from the OECD website: \url{http://www.oecd.org/pisa/pisaproducts/database-cbapisa2012.htm}.
Note that data presented in Table~\ref{tab:ticket} have been preprocessed from the raw PISA 2012 log file data and the variable names have been simplified.} In this example, the student was only aware of the fare of a concession daily ticket for city subway and purchased it.
He/she  did not check the fare of four concession individual tickets. Thus, although the ticket the student bought is a concession one and can be used for four trips by city subway in a day, it is not the cheapest one and thus does not completely satisfy the  task requirement.



\begin{table}
  \centering
  \begin{tabular}{l|l|llllll}
    \hline
StID&Time&Network&Fare&Ticket&Number&End\\
\hline
17&0&NULL&NULL&NULL&NULL&0\\
17&7.3&CITY SUBWAY&NULL&NULL&NULL&0\\
17&17.1&CITY SUBWAY&CONCESSION&NULL&NULL&0\\
17&27.1&CITY SUBWAY&CONCESSION&DAILY&NULL&0\\
17&32.9&NULL&NULL&NULL&NULL&1\\
    \hline
  \end{tabular}
  \caption{Log file data of a student solving the second task of the TICKETS unit. The columns ``StID" and ``Time" give the ID of the student and the time stamp of the action. The columns ``Network", ``Fare", ``Ticket", ``Number", and ``End" show the state of the student, as a result of the event history.}\label{tab:ticket}
\end{table}

\subsection{A Marked Point Process View}

We now provide a mathematical treatment of log file data, taking a marked point process framework. Consider a continuous-time domain $[0, \infty)$, with the task starting at time $t = 0$. Let $J$ be the number of event types, where each event type corresponds to a state of the system that can repeatedly occur. For the above TICKETS example, each state corresponds to a different screen of the task interface that can be represented by the last five columns of Table~\ref{tab:ticket}.
We define 21 states for the TICKETS task as given in the appendix. With well-defined event types,
log file data can be recorded by a double sequence $(\mathcal T, \mathcal Y) = ((T_n)_{n\geq 1}, (Y_n)_{n\geq 1})$, where
$T_n \in [0,\infty)$ is the time stamp of an event satisfying $T_n < T_{n+1}$,
and $Y_n \in \{1, 2, ..., J\}$ denotes the event types. Such a double sequence can be modeled by a marked point process \citep{cox1980point}, a stochastic process model commonly used in event history analysis \citep{cook2007statistical}.

A marked point process can be used to describe how future events depend on the event history at any time $t \in [0, \infty)$, where the event history is described by an information filtration $\mathcal F_t$. For log file data,
$\mathcal F_t = \{T_n, Y_n: T_n < t, n = 1, 2, ...\}$, which contains all available information up to time $t$.  A marked point process model can be characterized by a
\textit{ground intensity} function $\lambda(t\vert \mathcal F_t)$ and \textit{conditional density} functions $f(k\vert t, \mathcal F_t)$; see \cite{rasmussen2018lecture} for a review. In particular, the ground intensity function $\lambda(t\vert \mathcal F_t)$ describes the instantaneous probability of event occurrence, i.e.,
$$\lambda(t\vert \mathcal F_t) = \lim_{\Delta \rightarrow 0_+}\frac{P(T_{m+1} \in [t, t+\Delta)\vert \mathcal F_t)}{\Delta}, \mbox{~for~} m = \max\{n: T_n < t\}.$$
A task typically has a terminal state. Once the terminal state is reached, the task is completed and no event will happen afterwards, i.e.,
$\lambda(t\vert \mathcal F_t) = 0$, for $t$ greater than the time of reaching the terminal state. For the TICKETS example, the terminal state is reached, once a student clicks the ``BUY" button.

In addition, the conditional density function describes the instantaneous conditional probability of the $j$th type of event occurring, given that one event will occur, i.e.,
\begin{equation*}\label{eq:cond}
f(j \vert t, \mathcal F_t) = \lim_{\Delta \rightarrow 0_+} P(Y_{m+1} = j \vert \mathcal F_t, T_{m+1} \in [t,t+\Delta)), \mbox{~for~} m = \max\{n: T_n < t\}.
\end{equation*}
In our application, the conditional density functions often satisfy some zero constraints, because
some types of events cannot happen immediately
after some others. For the TICKETS task, such constraints are brought by the design of the system interface. For example, one cannot immediately reach the state (CITY SUBWAY, CONCESSION, NULL, NULL, 0) from the state (NULL, NULL, NULL, NULL, 0), where the five elements of a state correspond to the last five columns of Table~\ref{tab:ticket}.
We use $S({\mathcal F_t})$ to denote all the reachable states at time $t$ given event history $\mathcal F_t$. Then for any $j \notin S({\mathcal F_t})$, $f(j \vert t, \mathcal F_t) = 0$. 
For $j \in S({\mathcal F_t})$, the total probability law needs to be satisfied by the definition of conditional density functions, i.e.,
$$\sum_{j \in S({\mathcal F_t})} f(j \vert t, \mathcal F_t) = 1.$$

\begin{figure}
  \centering
  \includegraphics[scale = 0.7]{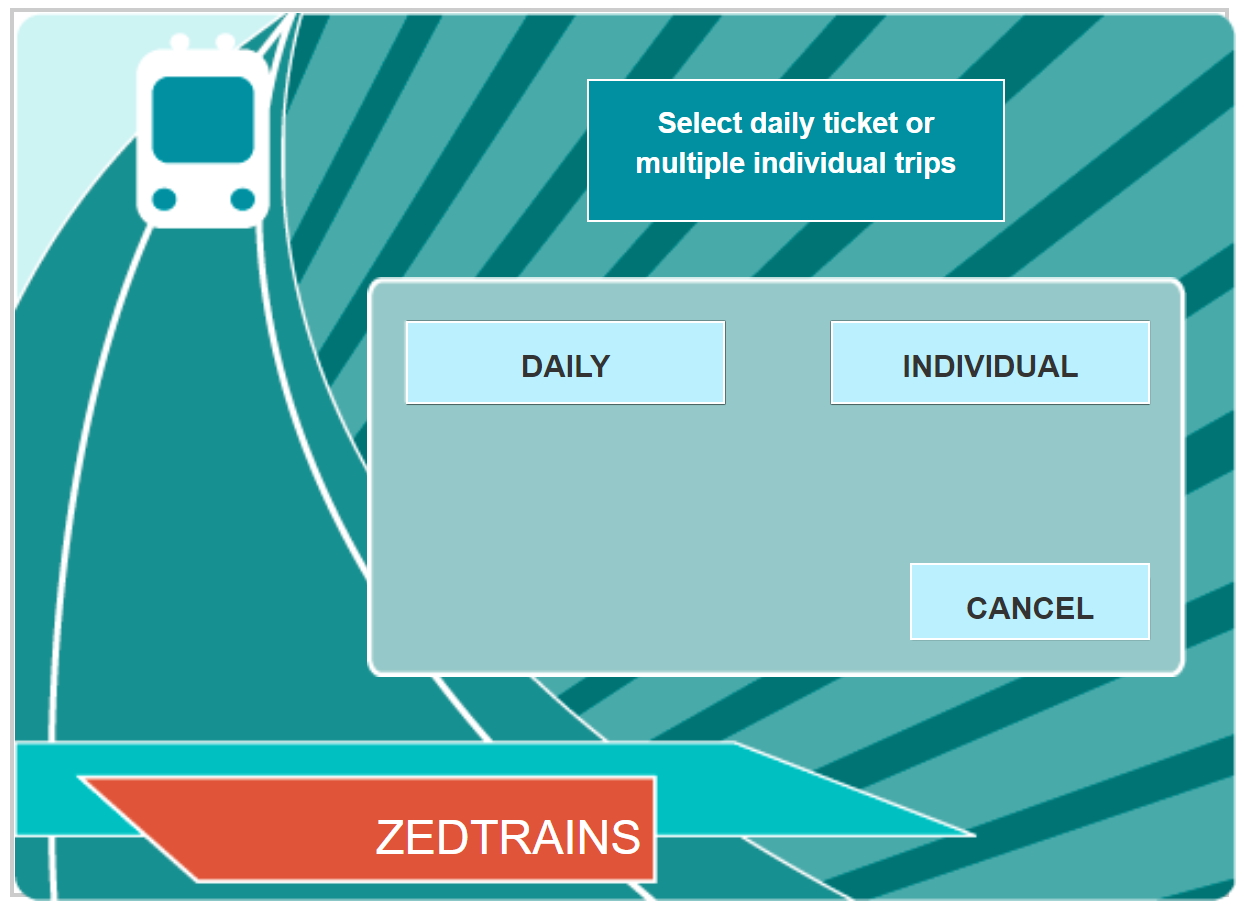}
  \caption{Screen shot of the system at state (CITY SUBWAY, CONCESSION, NULL, NULL, 0). }\label{fig:ticket3}
\end{figure}

 For each event type $j \in S({\mathcal F_t})$, there exists a measure of its effectiveness given by the structure of the problem-solving task, denoted by $V_j(\mathcal F_t)$.
A larger value of $V_j(\mathcal F_t)$ indicates higher effectiveness of event type $j$ as the next action. For the above TICKETS example, the
effectiveness of an action can be measured by whether it contributes to the final success of solving the task.
If an action contributes to the final success, then we set  $V_j(\mathcal F_t) = 1$, and otherwise $V_j(\mathcal F_t) = 0$.
For example, at the starting screen (see Figure~\ref{fig:ticket}), the action of clicking ``CITY SUBWAY" is always an effective action given the requirement of the task, while clicking ``COUNTRY TRAIN" or ``CANCEL" is not.   It is worth pointing out that whether or not an action is effective depends on the event history. Suppose that a student is currently at state (CITY SUBWAY, CONCESSION, NULL, NULL, 0), the screen of which is shown in Figure~\ref{fig:ticket3}. If neither the concession  fare of a daily subway ticket nor that of four individual subway tickets is known, then clicking either ``DAILY" or ``INDIVIDUAL" is effective but clicking CANCEL is not. However, if according to the event history the fare of a concession daily subway ticket is known
while that of four concession individual subway tickets is unknown,  then only clicking ``INDIVIDUAL" is effective at the current stage. A complete list of $V_j(\mathcal F_t)$ is shown in the appendix.

Data from $K$ tasks can be viewed as $K$ marked point processes. Thus, all the above quantities are task-specific and will be indexed by $k$.
Table~\ref{tab:keyele} summarizes the key elements for describing and modeling log file data from the $k$th task.  In what follows, we discuss the parametrization of the ground intensity and the conditional density functions, which links together  person-specific
latent traits, the structure of tasks, and log file process data.

\begin{table}
  \centering
  \small
  \begin{tabular}{c|l}
    \hline
    Notation & Interpretation\\
    \hline
    $Y_{kn}$ & The type (mark) of the $n$th event in the process of task $k$.\\
    $T_{kn}$ & The time stamp of the $n$th event in the process of task $k$. \\
    $\mathcal Y_k$ & $\mathcal Y_k = (Y_{kn})_{n\geq 1}$ denotes the  event sequence in the process of task $k$.  \\
    $\mathcal T_k$ &  $\mathcal T_k = (T_{kn})_{n\geq 1}$ denotes the  sequence of time stamps in the process of task $k$.  \\
    $\mathcal F_{kt}$ & The event history at time $t$ for task $k$, where $\mathcal F_{kt} = \{T_{kn}, Y_{kn}: T_{kn} < t, n = 1, 2, ...\}$.\\
    $S_{k}(\mathcal F_{kt})$& The set of event types that can immediately occur at time $t$ for task $k$. \\
    $V_{kj}(\mathcal F_{kt})$& The measure of effectiveness for event type $j$ of task $k$ at time $t$.\\
    $\lambda_k(t\vert \mathcal F_{kt})$& The ground intensity function of the marked point process for task $k$. It describes the \\
    &instantaneous probability of event occurrence. \\
    $f_k(j \vert t, \mathcal F_{kt})$ & The conditional density functions of the marked point process.
    It describes the  \\
    & instantaneous conditional probability of the $j$th type of event occurring for task $k$.  \\
    & $f_k(j \vert t, \mathcal F_{kt})=0$ for $j \notin S_{k}(\mathcal F_{kt})$.\\
    \hline
  \end{tabular}
  \caption{A list of the key elements for describing and modelling log file data.}\label{tab:keyele}
\end{table}

\section{Proposed Model}\label{sec:model}

\subsection{Specification of CTDC Model}\label{subsec:model}

We introduce two {continuous} person-specific latent variables, $\theta_i$ and $\tau_i$.
{As will be described below, these two latent variables will be used as parameters in a marked point process model to capture individual characteristics in problem-solving behaviors. More specifically, as will be discussed soon, $\theta_i$ and $\tau_i$ may be interpreted as student $i$'s problem-solving competency and action speed traits, respectively.  Like many other psychometric models with continuous latent variables,}
we assume $(\theta_i, \tau_i)$ to be bivariate normal, $N(\boldsymbol \mu, \Sigma)$, where $\boldsymbol\mu = (\mu_1, \mu_2)$ and $\Sigma = (\sigma_{ij})_{2\times 2}$.  

\begin{figure}
  \centering
  \includegraphics[scale = 1.5]{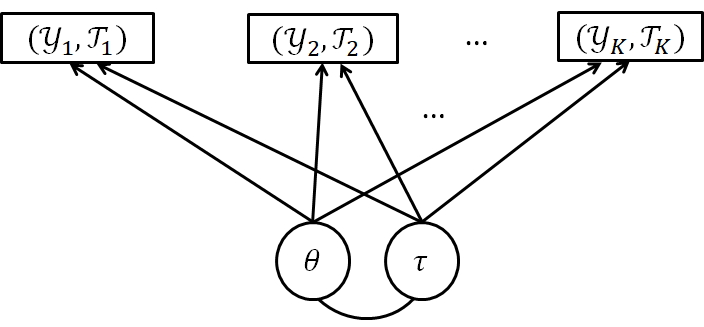}
  \caption{Path diagram for the proposed model, where $\theta$ and $\tau$ are the problem-solving competency trait and action speed trait, respectively,  and $(\mathcal{Y}_k, \mathcal{T}_k)$ denotes the log file process data from task $k$.}\label{fig:model}
\end{figure}

We consider log file process data from $K$ tasks that can be viewed as $K$ marked point processes.
We first assume local independence across tasks.
That is, we assume the $K$ marked point processes to be conditionally independent, given the two latent traits.
Figure~\ref{fig:model} provides the path diagram for the proposed model, where the details of the model
will be introduced in the sequel.


\begin{figure}
  \centering
  \includegraphics[scale = 1.5]{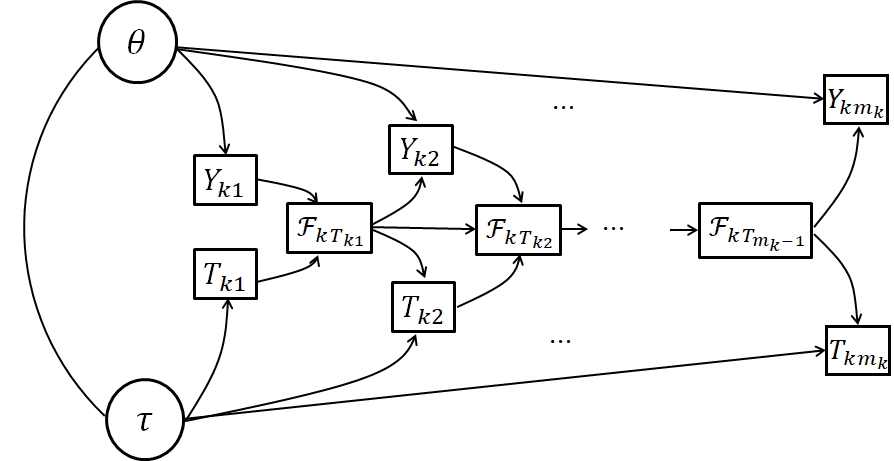}
  \caption{Path diagram for the proposed within-task model for each task $k$, where $\theta$ and $\tau$ are the problem-solving competency trait and action speed trait, respectively.}\label{fig:model2}
\end{figure}

Under the local independence assumption,
it suffices to model
data from one task. Specifically, we propose a model to describe how the conditional density functions and the ground intensity function depend on the two latent traits. Figure~\ref{fig:model2} provides the path diagram for the proposed within-task model. In this model, the next action, as modeled by the conditional density function,  depends only on the problem-solving competency trait and the event history. It does not directly depend on the action speed trait. In addition, the time stamp of the next action, as  modeled by the ground intensity function, depends only on the action speed factor and the event history. It does not directly depend on the competency trait.
The specifications of the submodel for actions and that for time stamps are described below, respectively.

%
%

\paragraph{Conditional density functions.} A conditional density function describes the conditional probability of
a student choosing state $j$ given that he/she will take an action in the next moment. It can be viewed as a discrete choice model.
Consider the conditional density function for event type $j$ of task $k$ at time $t$.
We adopt a multinomial logit model, taking the form
\begin{equation}\label{eq:bol}
f_k(j \vert t, \mathcal F_{kt}, \theta,  \beta_k) = \frac{\exp((\beta_k +  \theta)  V_{kj}(\mathcal F_{kt}) )}{\sum_{i\in  S_{k}(\mathcal F_{kt})}\exp((\beta_k + \theta) V_{ki}(\mathcal F_{kt}))}, \mbox{~for~} j \in   S_{k}(\mathcal F_{kt}),
\end{equation}
where $\beta_k$ is a task-specific easiness parameter and the rest of the notations are introduced previously in Table~\ref{tab:keyele}.
This choice model takes the form of a Boltzmann machine, which is similar to the within-task choice model in \cite{lamar2018markov}. It is a divide-by-total type model that is commonly used in the item response theory (IRT) literature \citep[e.g.,][]{thissen1986taxonomy}.

By the definition of $V_{kj}(\mathcal F_{kt})$ and given $\beta_k$, the larger the value of $\theta$, the more likely the effective actions will be taken.
In particular, when $\theta = \infty$, $f_k(j \vert t, \mathcal F_{kt}, \theta,  \beta_k) = 0$, for all $j$ such that  $ V_{kj}(\mathcal F_{kt}) \neq \max\{V_{ki}(\mathcal F_{kt}): i \in S_{k}(\mathcal F_{kt})\}$. That is, the most effective actions will be chosen with probability one.
Similarly,
when $\theta = -\infty$, $f_k(j \vert t, \mathcal F_{kt}, \theta,   \beta_k) = 0$, for all $j$ such that  $V_{kj}(\mathcal F_{kt})  \neq \min\{V_{ki}(\mathcal F_{kt}): i \in S_{k}(\mathcal F_{kt})\}$; i.e., the most ineffective actions will always be taken. Moreover, when $\beta_k + \theta = 0$, $f_k(j \vert t, \mathcal F_{kt}, \theta,  \beta_k) = {1}/{\vert S_{k}(\mathcal F_{kt}) \vert}$, for all $j \in S_{k}(\mathcal F_{kt})$.  In that case, the student performs in a purely random manner. {We emphasize that  $V_{kj}(\mathcal F_{kt})$ is a given effectiveness measure of the event type that depends on the problem-solving history. That is, whether  an action is effective or not at a given time point  depends on the actions that have been taken previously. See Section~\ref{subsec:exmp} for an example.}


In this action choice submodel \eqref{eq:bol},
parameter $\beta_k$ reflects the overall easiness of the task. Controlling for the value of $\theta$, tasks with a larger value of $\beta_k$ tend to be easier, as the effective actions are more likely to be chosen.






\paragraph{Ground intensity.} The ground intensity function essentially describes the speed of a student taking actions. For simplicity, we assume a student keeps a constant speed within a task once he/she has started working on the problem. That is,
\begin{equation}\label{eq:intensity}
\lambda_k(t\vert \mathcal F_{kt}, \tau,  \gamma_k) = \exp(\gamma_k +  \tau),
\end{equation}
for $\mathcal F_{kt}$ satisfying $T_{k1} < t$. An exponential form is assumed, as an intensity function has to be non-negative.
Here, $\gamma_k$ gives the baseline intensity of taking actions in solving task $k$.
The larger the $\gamma_k$, the faster the students proceed in general.
Given $\gamma_k$, the larger the value of $\tau$, the sooner the next action will be taken. In fact, it is easy to show that the expected time to the next action is $\exp(-\gamma_k - \tau)$.

We point out that
the first action needs to be treated differently, as the time to the first action involves not only taking an action, but also reading and understanding the requirement of the task.
In the proposed method, we do not specify a model for $T_{k1}$. Instead, all the inference will be based on a conditional likelihood estimator, in which $T_{k1}$ is conditioned upon.



\subsection{Inference}

\paragraph{Estimation.} We set the means of the latent traits $\mu_1 = \mu_2 = 0$ to ensure the identifiability of the
task-specific parameters. Thus, the fixed parameters of the model include $\beta_k$, $\gamma_k$, $k = 1, ..., K$, and $\Sigma$.
These parameters are estimated by a maximum marginal likelihood (MML) estimator.
Consider $N$ students taking the tasks.
We denote $(\mathcal T_{ik}, \mathcal Y_{ik})$ as the observed process data from student $i$ for task $k$, $i = 1, ..., N$, $k = 1, ..., K$, where $\mathcal T_{ik} = \{t_{ikn}: n = 1, ..., m_{ik}\}$ and
$\mathcal Y_{ik} = \{y_{ikn}: n = 1, ..., m_{ik}\}$, and $m_{ik}$ is the total number of actions taken by student $i$ on task $k$.
Recall that $\theta_i$ and $\tau_i$ are the latent traits of student $i$.

We derive the likelihood function based on the conditional distribution of $(\mathcal T_{ik}, \mathcal Y_{ik})$ given $T_{ik1}$, $\theta_i$, and $\tau_i$. This conditional likelihood function takes the form
\begin{equation*}
\begin{aligned}
L_{ik}(\ttt_i,  \beta_k,  \gamma_k)
=&\left(\prod_{n=1}^{m_{ik}} f_k(y_{ikn} \vert t_{ikn}, \mathcal F_{kt_{ikn}}, \theta_i,  \beta_k) \right)\\
 &\times  \left( \prod_{n=1}^{m_{ik} - 1} \exp\big(\gamma_k + \tau_i) \exp(-(t_{ik,n+1} - t_{ikn})\exp(\gamma_k + \tau_i)\big)\right),
\end{aligned}
\end{equation*}
where we denote $\ttt_i  = (\theta_i, \tau_i)$ to simplify the notation.
Making use of the across-task local independence assumption,
the marginal likelihood function takes the form

\begin{equation}\label{eq:mml}
l(\boldsymbol\beta, \boldsymbol\gamma, \Sigma) = \sum_{i=1}^N \log \left(\int
\prod_{k=1}^K L_{ik}(\ttt,  \beta_k,  \gamma_k ) \phi(\ttt| \Sigma) d\ttt \right),
\end{equation}
where $\phi(\cdot| \Sigma)$ is the probability density function of a bivariate normal distribution with mean $\mathbf 0$ and covariance matrix $\Sigma = (\sigma_{ij})_{2\times 2},$
and 
$\boldsymbol\beta = (\beta_1, ..., \beta_K)$, 
and $\boldsymbol\gamma = (\gamma_1, ..., \gamma_K)$.
Then our MML estimator of $(\boldsymbol\beta, \boldsymbol\gamma, \Sigma)$ is
\begin{equation}\label{eq:lik}
(\hat{\boldsymbol\beta}, \hat{\boldsymbol\gamma}, \hat{\Sigma}) = \argmax_{\boldsymbol\beta,  \boldsymbol\gamma, \Sigma} l(\boldsymbol\beta, \boldsymbol\gamma, \Sigma), \mbox{~s.t.~} \Sigma \succcurlyeq 0,
\end{equation}
where $\Sigma \succcurlyeq 0$ denotes the positive semi-definiteness of $\Sigma$.
The computation of \eqref{eq:lik} is carried out using an Expectation-Maximization (EM) algorithm \citep{dempster1977maximum}.
Given the estimated fixed parameters, the latent traits can be estimated using either the
expected a priori (EAP) estimator or the maximum a priori (MAP) estimator. In the subsequent analysis, the EAP estimator is adopted.






%

\subsection{Connections with Related Models}

In what follows, we make connections between the proposed model and related models in the psychometric literature.

\paragraph{Connection with MDP measurement model.}

We first compare the proposed model with  the MDP measurement model of  \cite{lamar2018markov}, which models the action sequence of a student solving a single task. Specifically, in \cite{lamar2018markov}, each student's action sequence is
described by a discrete-time MDP which also depends on a person-specific latent trait.
In this MDP,
the next action follows a choice model in a similar form as \eqref{eq:bol}, but the effectiveness measure $V_{kj}(\mathcal F_{kt})$ is replaced by the $Q$-function value of the process.
Given the MDP, the $Q$-function value can be obtained by solving an optimization problem.
As a result, there is no need to specify a measure of effectiveness for each possible action at any time point.
This feature makes the MDP measurement model very suitable for complex tasks that can be solved using many different strategies (e.g., board games), where the effectiveness of each potential action can be hard to specify.

However, the power of the MDP measurement model comes with a high computational cost,
as its estimation requires to iteratively alternate between updating person parameters and solving MDPs by dynamic programming.
For relatively simple tasks like the above TICKETS example and many other tasks used in large-scale assessments,
the action effectiveness can be reasonably specified. For such tasks, the proposed model is more suitable, given its dominant computational advantage.

Moreover, the proposed model makes use of information from both the action sequence and time stamps, 
while the MDP measurement model only focuses on the action sequence.
In particular, time stamps are incorporated into the proposed model through a continuous-time marked point process view of the log file data. {However, we also point out that, in order to model the time stamps,
the current model makes more assumptions  than the MDP measurement model. As a result, the proposed method may be more likely to suffer from model lack of fit, due to the potential misspecification of the submodel for time stamps.}

\paragraph{Connection with IRT models.} We make several connections between the proposed model and IRT models. First, the action choice submodel  \eqref{eq:bol} can be viewed as a nominal response model of a divide-by-total type \citep{thissen1986taxonomy}. Each action here is similar to an item in IRT.  The key difference is that the actions in the current model are not conditionally independent given the latent trait level, while such an conditional independence assumption is typically adopted for items in IRT models. In the proposed model, conditional dependence is introduced in a sequential manner, where the choice of an action can depend on the previous actions.
In addition, nominal response models in IRT typically have choice-specific parameters, while
the proposed model does not contain event-type-specific or event-history-specific parameters. This is because, the number of event types can be large, and the possible states of the event history can be even larger. Introducing such parameters can result in poor model performance due to the high variance in parameter estimation.

Second, the introduction of the competency and speed traits is similar in spirit to \cite{van2007hierarchical}'s joint model for item responses and response times. Specifically, \cite{van2007hierarchical} models the joint distribution of item-level responses and response times with two latent traits, one on  competency (i.e., ability) and the other on speed, respectively.
The item responses and response times in \cite{van2007hierarchical} are analogous to the actions and the time gaps between actions in our setting, respectively. Similarly,
in \cite{van2007hierarchical},
the item responses only depend on the competency trait and the response times only depend on the speed trait, and a correlation is allowed between the two latent traits.
In some sense, the proposed model can be viewed as an extension of \cite{van2007hierarchical} for process data, where the major difference is 
the introduction of event history in the current model to account for temperal dependence.


Third, the proposed model for data from multiple tasks induces an IRT model for the task outcomes. More precisely, we denote $Z_k$ as the final outcome for task $k$, where $Z_k = 1$ if the task is successfully solved and $Z_k = 0$, otherwise. Note that $Z_k$ is
a deterministic function of the action sequence $\mathcal Y_k$. As a result, based on the across-task local independence assumption and the
specification of the within-task model as given in Section~\ref{subsec:model}, $Z_1$, ..., $Z_K$ are conditionally independent given the competency trait $\theta$. Moreover, the probability $P(Z_k = 1\vert \theta)$ will be a monotone increasing function of $\theta$ under very
mild regularity conditions on the task structure; i.e.,  a higher the competency level leads to higher chance of solving the task. In that case, the final outcomes $Z_1$, ..., $Z_K$ given $\theta$ essentially follows a nonparametric monotone IRT model  \citep{ramsay1989binomial}.




\section{Case Study}\label{sec:real}

\subsection{Data}

To demonstrate the proposed CTDC model, we apply it to log file data from the first two tasks of the TICKETS unit in PISA 2012.
The TICKETS unit contains three tasks, among which the second task is introduced in Section~\ref{subsec:exmp} as a motivating example. In the first task, the students were asked to buy a full fare, country train ticket with two individual trips. This task is relatively simple.
To solve the task, one
first needs to select the network ``COUNTRY TRAINS", then choose the fare type ``FULL FARE", choose ticket type ``INDIVIDUAL",  select the number of tickets ``2", and finally click the ``BUY" button.

\begin{figure}
  \centering
  \includegraphics[scale = 0.5]{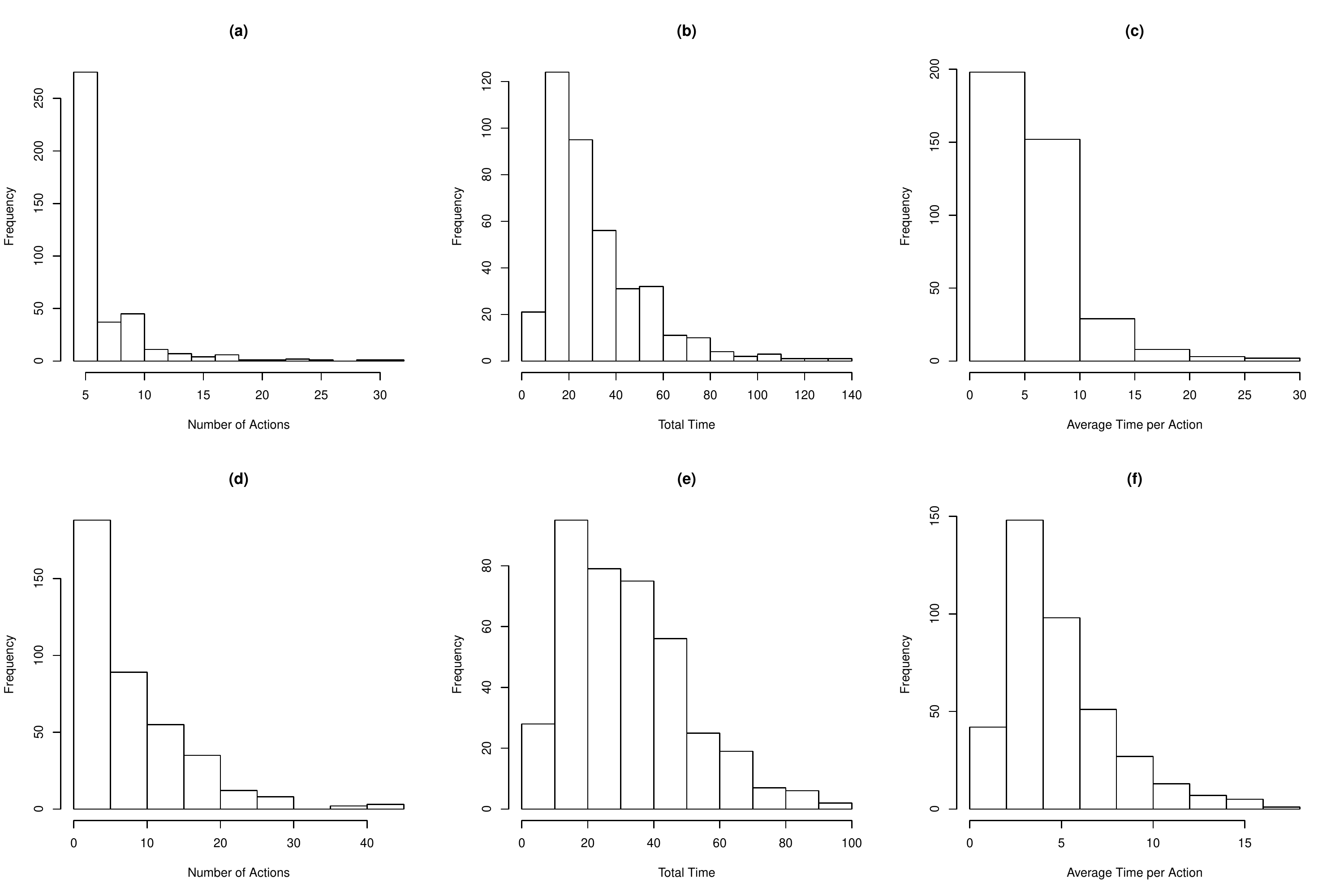}
  \caption{Histograms of summary statistics of process data. Data from the first task are visualized in
  panels (a)-(c) and data from the second task are visualized in panels (d)-(f). For each task, the three panels show the histograms for the number of actions, the total duration of problem solving, and the average time per action, respectively. }\label{fig:realsum}
\end{figure}

We analyze log file process data from the first two tasks of the unit\footnote{The first task of this Unit is available from \url{http://www.oecd.org/pisa/test-2012/testquestions/question4/}. The third task is not included in the analysis, as its
user interface is not publicly available though its description can be found in \cite{organisation2014pisa} and its data have been released.}. These data are from 392 United States students who completed both tasks. For simplicity, students who gave up in one of the two tasks during the problem-solving process are excluded from this analysis. The list of states and effectiveness of event types for the first task is given in the appendix.
Among the 392 students, 266 successfully solved the first task, 115  successfully solved the second, and 97 solved both. Figure~\ref{fig:realsum} 
shows the histograms of three summary statistics for the process data, including students' total number of actions, total duration, and average time per action.
Note that time to first action is included in calculating total duration, but is excluded when calculating average time per action.


The latent traits extracted by the proposed model will be validated
by comparing them with
the students' overall performance in PISA 2012 on problem-solving tasks. More precisely, PISA 2012 has in total  16 units of the problem-solving tasks. These 16 units
were grouped into four clusters, each of which was designed to be completed in 20 minutes.
Each student was given either one or two clusters.
Students' problem-solving performance was scaled using an IRT model based on the outcomes of the tasks they received \citep{OECD2014pisa}. For each student, five plausible values were generated from the corresponding posterior distribution of a proficiency  trait \citep{OECD2014pisa}. Following \cite{greiff2015computer},
we use the first plausible value as the continuous overall problem-solving performance score of the students.

\subsection{Results}

\begin{figure}
  \centering
  \begin{subfigure}[t]{0.33\textwidth}
        \centering
        \includegraphics[scale = 0.5]{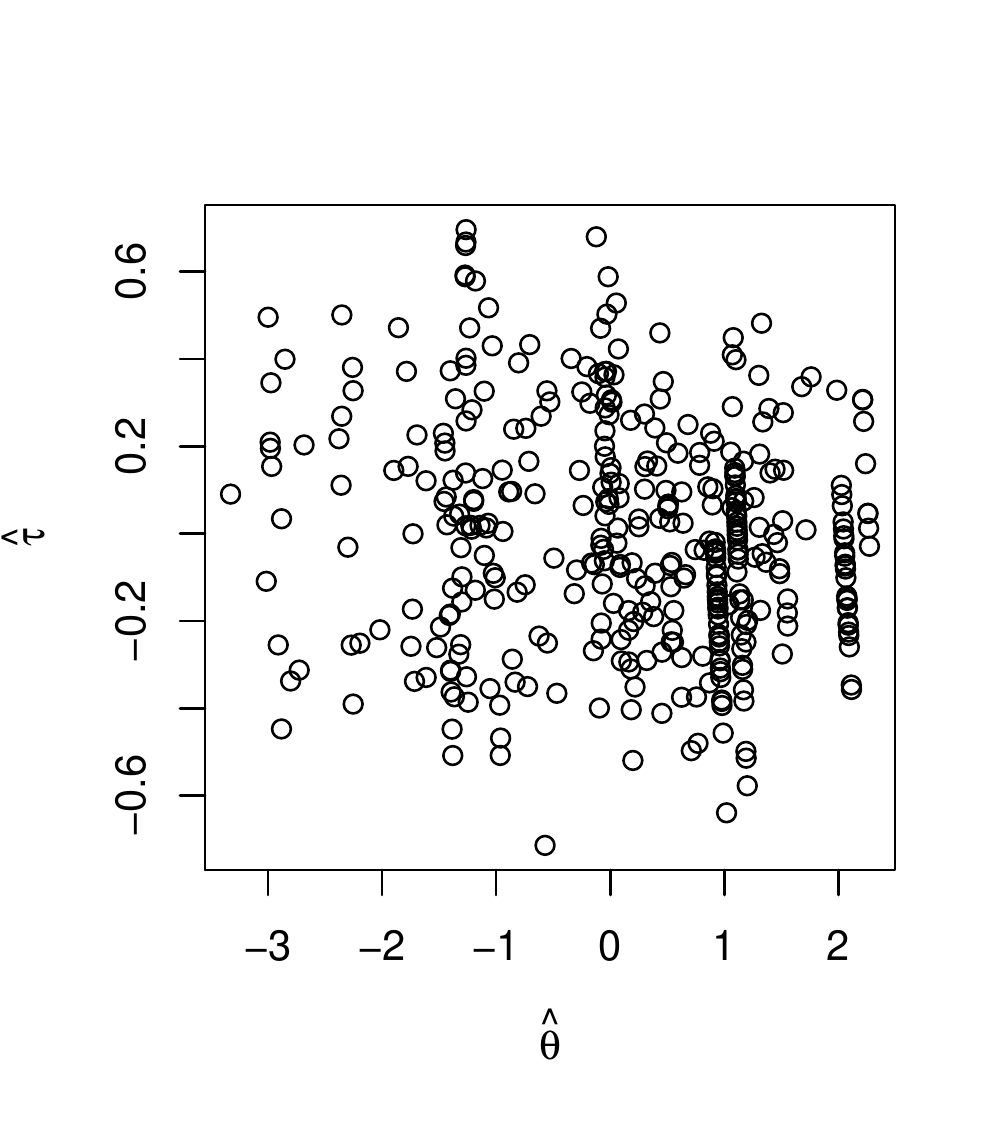}
        \caption{}
    \end{subfigure}%
    \begin{subfigure}[t]{0.33\textwidth}
        \centering
        \includegraphics[scale = 0.5]{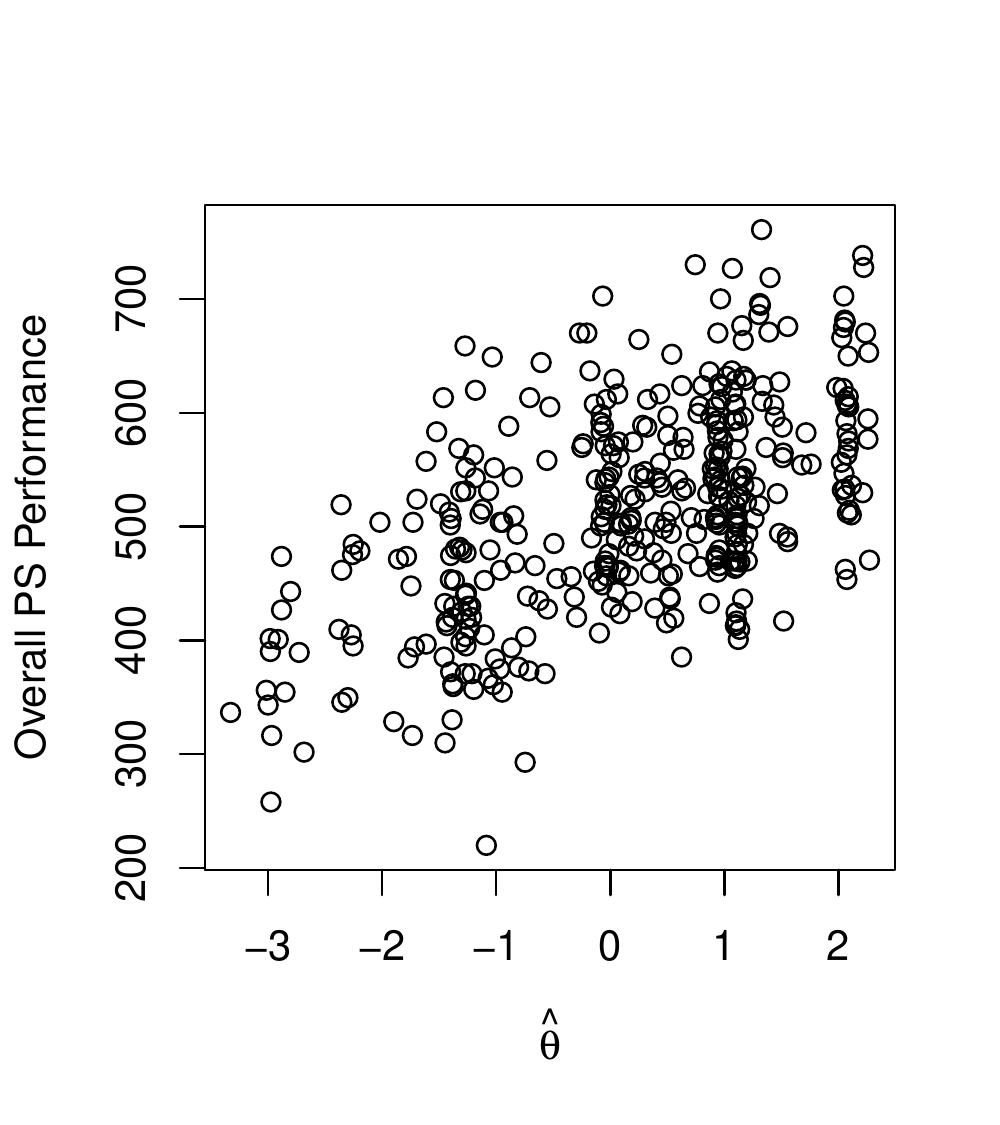}
        \caption{}
    \end{subfigure}
    \begin{subfigure}[t]{0.33\textwidth}
        \centering
        \includegraphics[scale = 0.5]{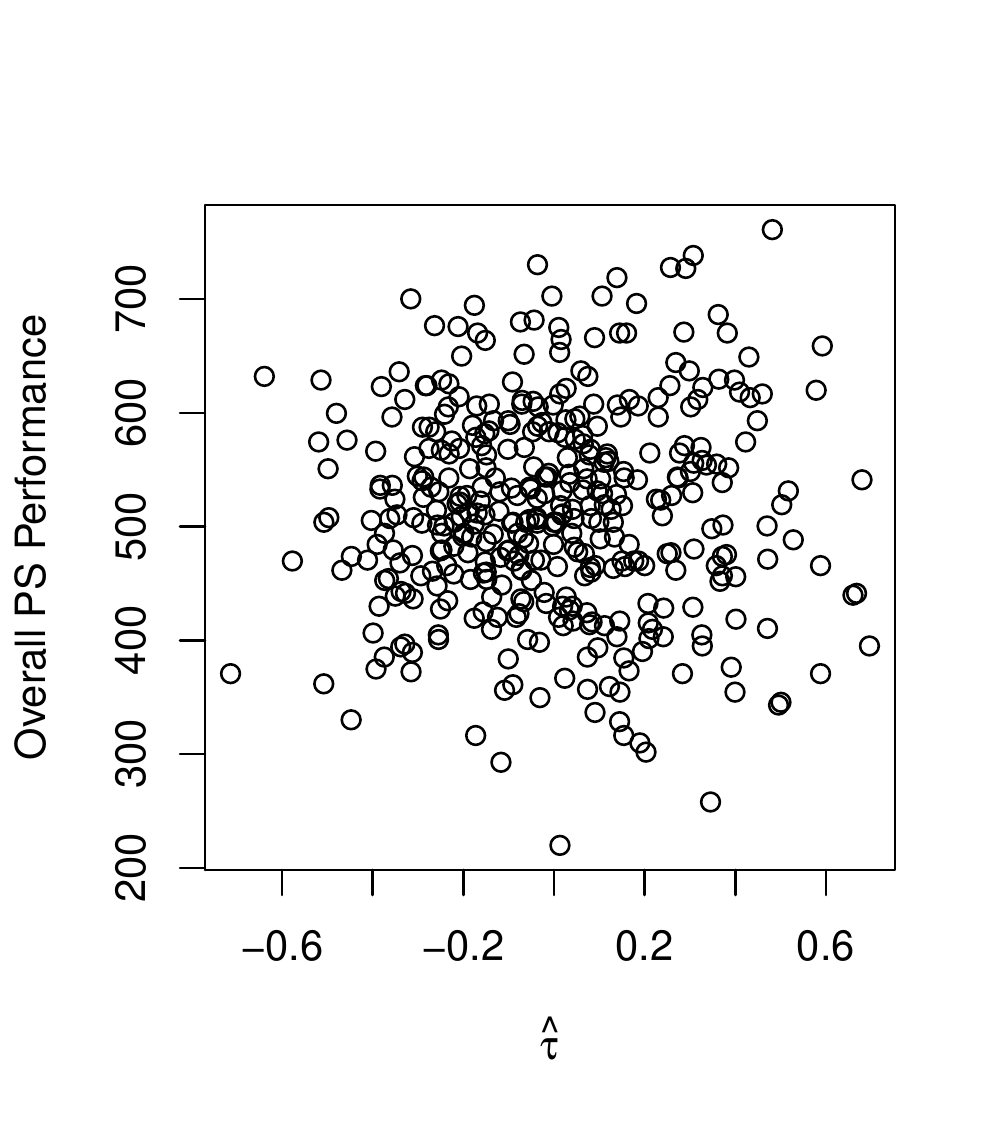}
        \caption{}
    \end{subfigure}
  \caption{Real data analysis. Panel (a): The scatter plot of the EAP estimates of the two latent traits. Panel (b): The scatter plot of the EAP estimate of the problem-solving competency trait (x-axis) versus the overall performance score (y-axis). Panel (c): The scatter plot of the EAP estimate of the action speed trait (x-axis) versus the overall performance score (y-axis).}\label{fig:EAP}
\end{figure}

\paragraph{Parameter estimation.} We apply the proposed model to data from the two tasks. The MML estimate of the fixed parameters is given in Table~\ref{tab:real.est1}.
The estimated correlation between the two latent traits is $\hat \sigma_{12}/\sqrt{(\hat{\sigma}_{11}\hat{\sigma}_{22})} = -0.11$, with a $95\%$ confidence interval $(-0.26, 0.04)$. It suggests that the problem-solving competency trait and the action speed trait
have a very weak negative correlation that is not significantly different from zero.
Panel (a) of Figure~\ref{fig:EAP} provides the scatter plot of the EAP estimates of the two latent traits, where no clear association can be found between the estimated traits.


\begin{table}
  \centering
  \begin{tabular}{lcccccccccccc}
    \hline
    Parameter & $\beta_1$ & $\beta_2$  & $\gamma_1$ & $\gamma_2$  & $\sigma_{11}$ & $\sigma_{12}$ & $\sigma_{22}$ \\
    \hline
    MMLE & 1.54 & 1.68 & -1.73 & -1.37 & 2.18 & -0.06 & 0.11  \\
    \hline
    SE & 0.08 &0.08& 0.03& 0.03& 0.22& 0.04& 0.01 \\
    \hline
  \end{tabular}
  \caption{Real data analysis: MML estimates of the fixed parameters and their standard errors.}\label{tab:real.est1}
\end{table}

According to the estimated easiness parameters $\beta_1$ and $\beta_2$ as shown in Table~\ref{tab:real.est1},
the second task is slightly easier in the choice of effective actions within a task, though the second task seems more difficult according to its design and has a lower success rate according to the task outcome data. There are two possible explanations. First, the difficulty level of the first task may be boosted as it was the students' first encounter with this ticketing machine,
while in the second task, the students already had a good understanding of the system. This difference in the familiarity with the task interface can be reflected by the task-specific easiness parameters.
Second, although it is difficult for students to completely solve the second task, it is not very difficult to partially fulfill the requirements. That is, a student may purchase a daily subway ticket or four individual subway tickets in concession fare without comparing their prices.
In this process, many effective actions are taken, which reduces the overall difficulty of the task.
Based on the estimated baseline intensities $\gamma_1$ and $\gamma_2$, the students tend to act slightly faster in the second task than in the first. This is possibly due to the students' increased level of familiarity with the task interface when solving the second task.

\paragraph{Validating the latent traits.}
We now investigate the relationship between the EAP estimates of the latent traits and student's overall performance score given by OECD. Panels (b) and (c) of Figure~\ref{fig:EAP} show the scatter plots of the EAP estimates of the two latent traits versus the overall performance score, respectively.
From these plots, a moderate positive association seems to exist between the estimated competency trait and the overall performance, while there seems no clear association between the estimated speed trait and the overall performance.

We further regress the overall performance score on the estimated traits to investigate their relationship.
Specifically, three models are fitted, denoted as models $\mathcal M_1$ through $\mathcal M_3$, respectively.
In these three models,
we regress the overall performance score on the estimated competency trait, the estimated speed trait, and both, respectively.
The parameter estimation results of these three models
are given  in Table~\ref{tab:real.Reg} and the
 $R^2$ values  are given in Table~\ref{tab:real.Rsq}.
According to the results of models $\mathcal M_1$ and $\mathcal M_3$, the competency trait extracted from the process data
is a significant predictor of the overall performance score. In particular, its slope parameter is positive in both models, meaning that students with a higher competency score tend to have better overall performance in problem solving.
In addition, based on the $R^2$ of model $\mathcal M_1$, the competency trait alone explains 32.34\% of the information in the overall performance score.

According to the result of model $\mathcal M_2$,
the speed trait alone has almost no explanation power of the overall performance, with its slope parameter insignificant ($p = 0.69$)
and $R^2$ value as small as $0.04\%$. Interestingly, however, the speed trait becomes significant ($p = 0.01$) in model $\mathcal M_3$ when both traits are included as covariates. Comparing with model $\mathcal M_1$, the increase in the $R^2$ value is 1.09\%, with a 95\% bootstrap confidence interval (0.03\%, 3.44\%).
 The slope estimate for the speed trait is positive, meaning that students with higher speed tend to have better overall performance, when controlling for their competency trait level.


\begin{table}[H]
  \centering
  \begin{tabular}{l|rrrrrcccccc}
    \hline
    \hline
    $\mathcal M_1$ &Estimate & SE & p-value (two-sided)   \\
    \hline
    Intercept &  508.44 &  3.82     &  $<2\times 10^{-16}$\\
    Slope (C) &   40.15 &  2.94     &  $<2\times 10^{-16}$    \\
    \hline
    \hline
    $\mathcal M_2$ &Estimate & SE & p-value (two-sided)   \\
    \hline
    Intercept&514.33 &     4.61 & $<2\times 10^{-16}$  \\
    Slope (S)&7.12   &     17.85  &    0.69 \\
    \hline
    \hline
    $\mathcal M_3$ &Estimate & SE & p-value (two-sided)   \\
    \hline
    Intercept&    508.45  &    3.79       &  $<2\times 10^{-16}$ \\
    Slope (C)&    41.23   &    2.95      &  $<2\times 10^{-16}$  \\
    Slope (S)&    37.11   &    14.74   &  0.01\\
    \hline
  \end{tabular}
  \caption{Real data analysis: The parameter  estimation results of three regression models which regress the overall performance score on
  the EAP estimate of the competency trait ($\mathcal M_1$), that of the speed trait ($\mathcal M_2$), and both ($\mathcal M_3$).}\label{tab:real.Reg}
\end{table}

\begin{table}[H]
  \centering
  \begin{tabular}{l|ccc|cc|ccccccc}
    \hline
    Model & $\mathcal M_1$ &$\mathcal M_2$ &$\mathcal M_3$ & $\mathcal M_4$ & $\mathcal M_5$ & $\mathcal M_6$ & $\mathcal M_7$ & $\mathcal M_8$\\
    \hline
    $R^2$&  32.34\% & 0.04\% & 33.43\% & 24.18\%  & 23.78\%  & 24.37\% & 17.47\% & 34.06\%   \\
    \hline
  \end{tabular}
  \caption{Real data analysis: The $R^2$ values of eight linear regression models, each of which takes the overall problem-solving performance score as the response variable. }\label{tab:real.Rsq}
\end{table}

\paragraph{Fitting CTDC model to single tasks.}
We further investigate the explanation power of the latent traits extracted from each single task.
That is,  we fit the proposed model to data from each single task and obtain the EAP estimate of the two traits.  Then we regress the overall performance score on the estimated traits.
This results in two regression models, denoted as $\mathcal M_4$ and $\mathcal M_5$, for the two tasks, respectively.
As given in Table~\ref{tab:real.Rsq},
the $R^2$ values of these models are 24.18\% and 23.78\%, respectively. Comparing model $\mathcal M_3$ with model $\mathcal M_4$, the improvement in the $R^2$ value is 9.25\%, with 95\% bootstrap confidence interval (4.46\%,  13.85\%). In addition, comparing model $\mathcal M_3$ with model $\mathcal M_5$, the improvement in $R^2$ is 9.65\%, with 95\% bootstrap confidence interval (3.56\%,  15.30\%). This result implies that the
joint analysis of the two tasks extracts more meaningful information than that of each single task. The information gain from adding one task in the analysis reflects its  unique information that is not shared with the other task.

%
%
%
%


\paragraph{Process data versus final outcome.}
We compare the explanation power of the extracted latent traits from the fitted models with those of the final outcomes.
{We are interested in whether process data contain more information about the students' ability than the final outcomes. More precisely, we define a student's binary final outcome of a task as whether he/she completely solves the task, as determined by the problem-solving process. For the first task, the final outcome is success if the student purchases a full fare, country train ticket with two individual trips. For the second task, the final outcome is success if the student purchases four individual subway tickets in concession fare after
comparing its price to that of a daily subway ticket in concession fare.}
Specifically,
in models $\mathcal M_6$ and $\mathcal M_7$, we regress the overall performance score on the binary final outcome (success/failure) of each single task, respectively. In model $\mathcal M_8$, we regress the overall performance on the outcomes of the two tasks.
The $R^2$ values of the fitted models are given in Table~\ref{tab:real.Rsq}.

First, we compare the $R^2$ values of models $\mathcal M_4$ and $\mathcal M_6$. Their difference is -0.19\%, with  95\% bootstrap confidence interval (-4.76\%,  3.78\%). This result implies that the process data of the first task may not provide more information than the final outcome. This is
not surprising, given that the requirement of the task is straightforward and the task can be solved using a small number of steps.

Second, we compare the $R^2$ values of models $\mathcal M_5$ and $\mathcal M_7$. Their difference is 6.31\%, with  95\% bootstrap confidence interval (0.11\%,  13.60\%). This result suggests that the process data from the second task seem to contain more information about the students' overall performance than the corresponding binary outcome.  This is likely due to that the second task is more complex. 

Finally, the difference in the $R^2$ values of models $\mathcal M_3$ and $\mathcal M_8$ is  -0.63\%, with  95\% bootstrap confidence interval
(-6.18\%, 5.32\%). It suggests that the process data of the two tasks do not contain significantly more information about the students' overall problem-solving performance than their final outcomes. This is likely due to that the information gain from the process data of the second task can be almost completely explained by the unique information in the first task.

\paragraph{Discussion.} We end this section with some discussions. First, the comparison based on the overall performance score may not be
completely fair. 
The information in the task final outcomes may be overestimated, due to the use of the  overall performance score as the standard for the way it is constructed.
It may be more fair to validate the extracted latent traits by the students' overall performance on the tasks
excluding the current ones.



Second, we point out that the amount of additional information process data contain is largely determined by the design of the tasks. We believe that tasks which are more complex and require more steps to solve  have more additional information in the process data.
For such tasks, we may only need a small number of tasks to accurately evaluate students' performance, by extracting information from process data.

Finally, the proposed model allows us to investigate task-specific characteristics of problem-solving processes,
including the difficulty level and the baseline intensity. Such information can provide useful feedbacks to the design of the tasks.



%


\section{Simulation}\label{sec:sim}

We now provide a simulation study to further investigate the proposed model and its estimation.

\paragraph{Simulation setting.} Following the setting of the real data example, we simulate data from two tasks (i.e., $K = 2$).  Two sample sizes are considered, including $N = 100$ and $400$. In addition, we consider three settings for the correlation between the two latent traits, including $\rho = -0.25, 0,$ and 0.25. The structure of the two tasks is set the same as those of the case study, and the model parameters except for $\sigma_{12}$ are set the same as the estimates in Table~\ref{tab:real.est1}. The covariance between the two traits $\sigma_{12}$ is determined by the correlation and the variances of the two traits, i.e., $\sigma_{12} = \rho \sqrt{\sigma_{11} \sigma_{22}}$.
This leads to six different settings, as listed in Table~\ref{tab:setting}. For each setting, we generate 50 independent replications using the proposed CTDC model.

\begin{table}
  \centering
  \begin{tabular}{l|ccc|ccc}
    \hline
    Setting& $S_1$& $S_2$& $S_3$& $S_4$& $S_5$& $S_6$\\
    \hline
    $N$ &  100&100&100& 400&400&400\\
    $\sigma_{12}$ & -0.25 & 0 & 0.25   & -0.25 & 0 & 0.25  \\
    \hline
  \end{tabular}
  \caption{Simulation study: The list of six simulation settings.}\label{tab:setting}
\end{table}

\begin{figure}
  \centering
  \includegraphics[scale = 0.4]{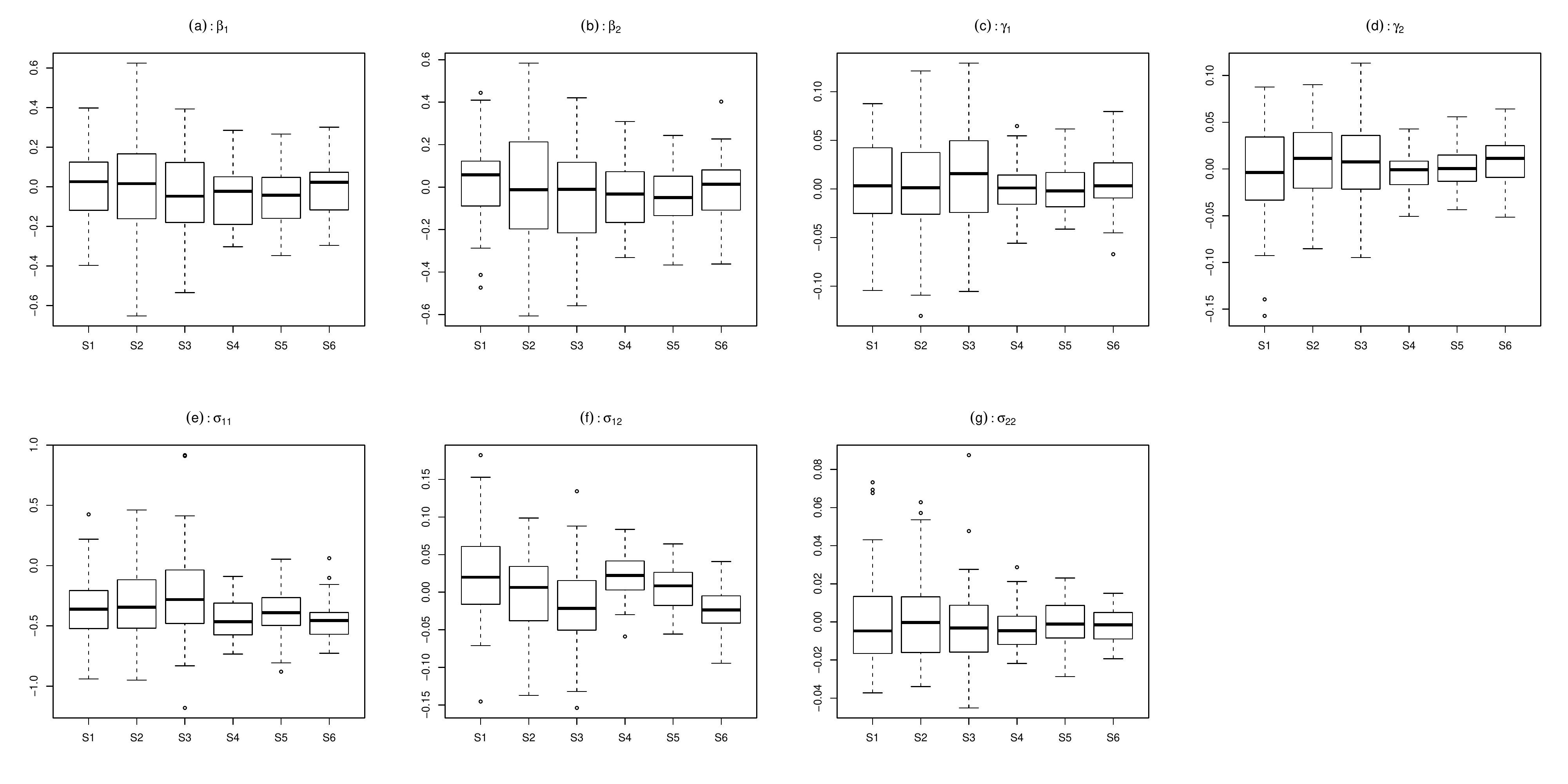}
  \caption{Simulation study: Estimation error of the seven fixed parameters. Panels (a)-(g) correspond to parameters $\beta_1$, $\beta_2$, $\gamma_1$, $\gamma_2$, $\sigma_{11}$, $\sigma_{12}$, $\sigma_{22}$, respectively.}\label{fig:bias}
\end{figure}

\paragraph{Results.} The estimation of the fixed parameters is shown in Figure~\ref{fig:bias}, where each panel corresponds to
a fixed parameter. In each panel,
six boxplots are shown that correspond 
to different simulation settings, respectively. Each boxplot shows the
estimation error of the corresponding parameter over 50 replications. As we can see, the MML estimate of the fixed parameters
is reasonably accurate under all the simulation settings. In addition, the estimation accuracy improves when the sample size increases. Moreover, the different settings for $\rho$ do not substantially affect the estimation accuracy of $\beta_k$s and $\gamma_k$s, but they do seem to affect the estimation accuracy for $\Sigma$.


\begin{figure}[h]
  \centering
  \includegraphics[scale = .45]{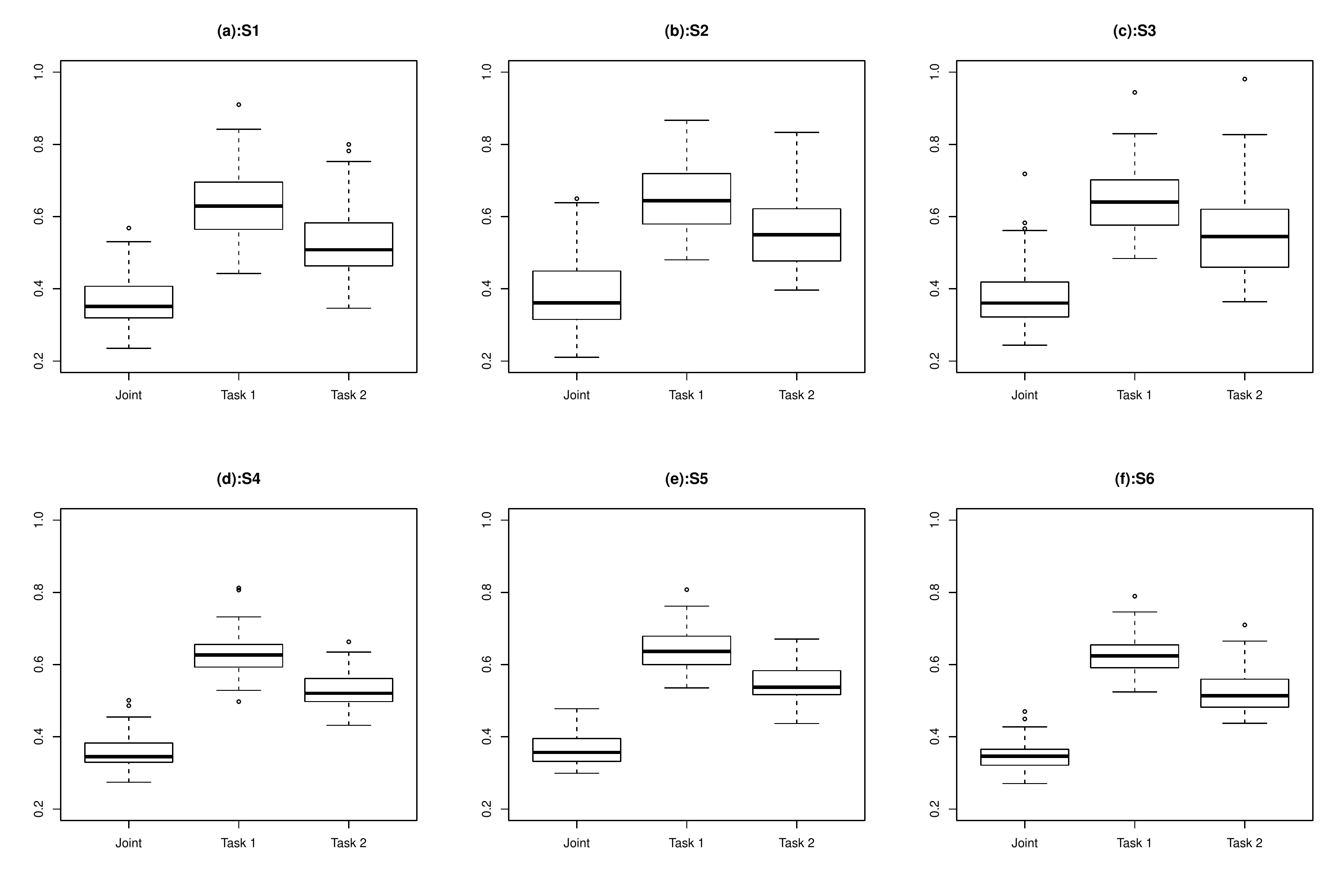}
  \caption{Simulation study: The mean squared error in the EAP estimate of the competency trait. The six panels correspond to the six simulation settings. In each panel, the three boxplots correspond to results based on (1) the joint analysis of the two tasks, (2) analysis of the first task, and (3) analysis of the second task, respectively. }\label{fig:traits}
\end{figure}

\begin{figure}[h]
  \centering
  \includegraphics[scale = .45]{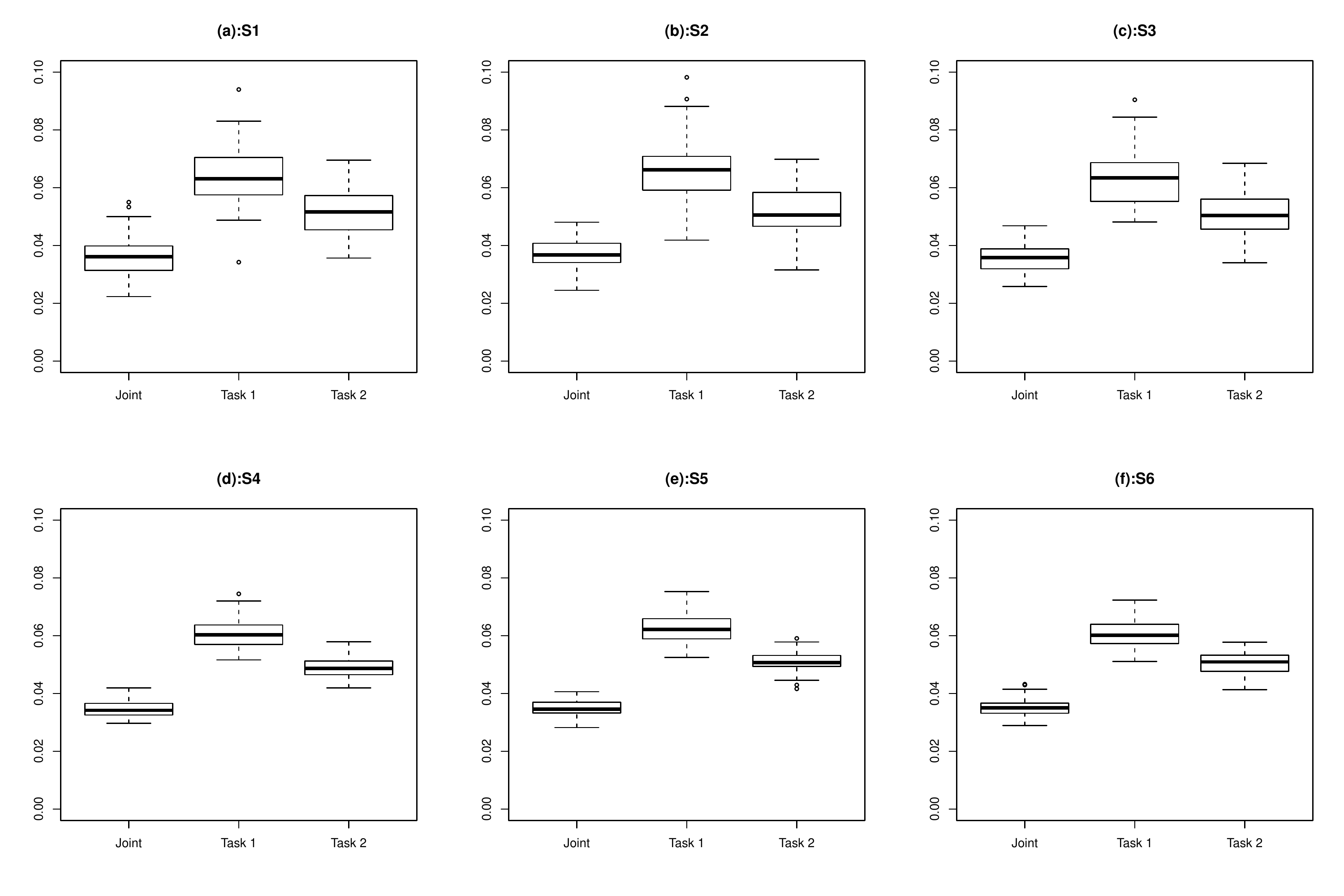}
  \caption{Simulation study: The mean squared error in the EAP estimate of the speed trait. The six panels correspond to the six simulation settings. In each panel, the three boxplots correspond to results based on (1) the joint analysis of the two tasks, (2) analysis of the first task, and (3) analysis of the second task, respectively. }\label{fig:traits2}
\end{figure}

We further look at the estimation of the latent traits.
Specifically, we measure estimation accuracy by the mean squared error (MSE) of the EAP estimate of the two traits.
The results are given in Figures~\ref{fig:traits} and \ref{fig:traits2}, where the two figures provide the results for the competency and speed traits, respectively.
In each figure, the six panels correspond to the six simulation settings, respectively.
For each panel of each figure, three boxplots are shown, where
the EAP estimate of the corresponding latent trait is based on (1) the joint analysis of the two tasks, (2) the first task, and (3) the second task, respectively. By comparing the first boxplot with the other two, we see that the joint analysis of the two tasks leads to
a higher accuracy in the estimation of the latent traits.
In addition, by comparing the second boxplot with the third, we see that data from the second task lead to more accurate estimation of the latent traits, suggesting that the second task tends to be more informative.
Furthermore, the between-replication variability tends to be smaller when the sample size becomes larger.



\section{Discussions}\label{sec:diss}

In this paper, we propose a latent variable model for measuring problem-solving related traits based on log file process data.
We take an event history analysis framework, under which data within a task
are modeled as a marked point process and then multiple tasks are linked together using
a local independence assumption. In the proposed model, a marked point process is characterized by two components, including (1) conditional density functions for sequential actions and (2) a ground intensity function for time stamps.
A parametrization of these two components is given that links together person-specific latent traits,  the structure of problem-solving
task, and log file  process data.
In particular, we model the conditional density functions using a Boltzmann machine choice model, where the chance of an action being chosen depends on the event history, the level of problem-solving competency trait, and a task-specific easiness parameter.
In addition, the ground intensity is assumed to depend on an action speed trait and a task-specific baseline intensity parameter.
The proposed model is applied to process data from two problem-solving tasks in PISA 2012. 
The estimated model parameters provide sensible characterizations of the tasks and the distribution of the two latent traits. The extracted latent traits are validated by comparing them with students' overall problem-solving performance score reported by PISA 2012. The main findings include: 
(1) both latent traits are significant predictors of students' overall performance, with the prediction power mainly from the competency trait, (2) the joint analysis of the two tasks provide more information than the analysis of each single task, and (3) the process data of the second task provide more information than its final outcome, while the process data of the first task does not seem to contain additional information.


%
%

We point out that the proposed method is very flexible in analyzing log file process data with different types of data missingness.
First of all, thanks to the across-task local independence assumption, the proposed method still applies when some students' data are missing completely at random (MCAR) on a subset of tasks (e.g., due to a planned missing data design). This is similar to
treating MCAR item responses in a local independence IRT model. Second, the proposed method is also powerful in handling data that are right censored in time within a task. More precisely,
a process is said to be right censored at time $t$, when data after time $t$ are not observed.  For example, right censoring can happen when a student does not have enough time to complete the task. Thanks to the statistical properties of marked point process, the proposed inference procedures can be easily extended to process data with an independent censoring time \citep[e.g.,][]{andersen1988censoring}. Thus, we can make statistical inference for students who do not complete one or multiple tasks.
{With the aforementioned advantages, the proposed model may be useful in low-stake tests, such as large-scale assessments, for the measurement and comparison of students' problem-solving ability. For example, this approach can be used to compare students who receive different tasks which may not be equally difficult. The model can also be used to assign students partial credits based on their problem-solving processes, when they fail to solve a task.}

{Despite its advantages as a measurement model, the proposed approach has some limitations. Specifically, the submodels for action choices and time stamps may be over-simplified, relative to the high-complexity of students' problem-solving behaviors. In fact, the proposed model may be more likely to suffer from model misspecification than the MDP measurement model, due to making additional assumptions in its submodel for time stamps. Thus, it is important to assess
the goodness-of-fit of the proposed model  when applying it to real tests. Assessing the goodness-of-fit of marked point process models is non-trivial, for which performance metrics and statistical theory remain to be developed. Given the potential model misspecification problem, we do not recommend to use it in the scoring of high-stake tests, i.e.,  tests with important consequences for the
students.}

{Complex tasks are often needed to measure problem-solving constructs that are hard to measure.
The proposed method requires to know a priori the effectiveness of any action given any possible event history. This requirement may limit its direct application to complex tasks for which the specification of action effectiveness is challenging. To extend the proposed method to analyzing more complex tasks, stochastic models and computational methods remain to be developed to automatically obtain a meaningful effectiveness measure based on the design of a task. }



We further discuss several future directions of the proposed method.
First, computationally more efficient methods may be developed for the estimation of the proposed model. 
Due to the complexity and size of process data and the numerical integrations involved, the EM algorithm adopted here may not be
sufficiently fast. In fact, computationally more efficient algorithms can be developed for the proposed MML estimator, such as the Metropolis-Hastings Robins-Monro algorithm \citep{cai2010high} and the stochastic EM algorithms \citep{celeux1985sem,diebolt1996stochastic,zhang2020improved}.
In addition, the joint likelihood estimator may be a good alternative estimator that treats the person-specific latent traits as fixed parameters \citep{haberman1977maximum,chen2019joint,chen2019structured}. 
Its computation is much faster than the MML estimator, as it avoids numerical or Monte Carlo integrations that is computationally intensive. Given the large amount of information for each student from process data, 
consistent estimation of both fixed parameters and latent variables may still be obtained.


Second,
similar to other latent-variable-based measurement models,
the proposed model can be combined with structural models to study the relationship between the problem-solving traits and other variables under a structural equation modeling framework. For example, for PISA data, it is often of interest to understand the relationship between students' problem-solving traits, and other variables including cognitive abilities and other background variables from the student, parent, and school questionnaires of the PISA survey.

Finally, this model can be extended to measure multiple latent traits, provided that design information is available about the traits needed in each step that may depend on the problem-solving event history. In fact, problem-solving behavior is likely driven by multiple latent traits. For example, the PISA 2012 framework
decomposes problem solving into four dimensions based on the corresponding cognitive processes, including ``exploring and understanding", ``representing and formulating", ``planning and executing", and ``monitoring and reflecting" \citep{organisation2014pisa}.
The current model can be extended to measure these finer-grained dimensions, when design information is available on the dimensional structure in each problem-solving step.


%
%



\bibliographystyle{apacite}
\bibliography{review}

\clearpage
\section*{Appendix}

\begin{table}[H]
\footnotesize
  \centering
  \begin{tabular}{cccccccc}
    \hline
State ID &Network&Fare&Ticket&Number&End \\
\hline
1&NULL&NULL&NULL&NULL&0\\
2&CITY SUBWAY&NULL&NULL&NULL&0\\
3&CITY SUBWAY&FULL&NULL&NULL&0\\
4&CITY SUBWAY&FULL&DAILY&NULL&0\\
5&CITY SUBWAY&FULL&INDIVIDUAL&NULL&0\\
6&CITY SUBWAY&FULL&INDIVIDUAL&1/2/3/4/5&0\\
7&CITY SUBWAY&CONCESSION&NULL&NULL&0\\
8&CITY SUBWAY&CONCESSION&DAILY&NULL&0\\
9&CITY SUBWAY&CONCESSION&INDIVIDUAL&NULL&0\\
10&CITY SUBWAY&CONCESSION&INDIVIDUAL&1/2/3/5&0\\
11&CITY SUBWAY&CONCESSION&INDIVIDUAL&4&0\\
12&COUNTRY TRAIN&NULL&NULL&NULL&0\\
13&COUNTRY TRAIN&FULL&NULL&NULL&0\\
14&COUNTRY TRAIN&FULL&DAILY&NULL&0\\
15&COUNTRY TRAIN&FULL&INDIVIDUAL&NULL&0\\
16&COUNTRY TRAIN&FULL&INDIVIDUAL&1/2/3/4/5&0\\
17&COUNTRY TRAIN&CONCESSION&NULL&NULL&0\\
18&COUNTRY TRAIN&CONCESSION&DAILY&NULL&0\\
19&COUNTRY TRAIN&CONCESSION&INDIVIDUAL&NULL&0\\
20&COUNTRY TRAIN&CONCESSION&INDIVIDUAL&1/2/3/4/5&0\\
21&NULL&NULL&NULL&NULL&1\\
\hline
  \end{tabular}
  \caption{A complete list of the 21 states of the second task of the TICKETS unit.}\label{tab:tickets2}
\end{table}

\begin{table}[H]
\footnotesize
  \centering
  \begin{tabular}{c|c|c|c|c|c|c|c|c}
    \cline{2-9}
 & \multicolumn{2}{c|}{A}&\multicolumn{2}{c|}{B}&\multicolumn{2}{c|}{C}&\multicolumn{2}{c}{D}\\
    \cline{2-9}
State&  \multicolumn{1}{|c|}{$+$} & \multicolumn{1}{c|}{$-$} &\multicolumn{1}{c|}{$+$} &\multicolumn{1}{c|}{$-$}& \multicolumn{1}{c|}{$+$} & \multicolumn{1}{c|}{$-$} &\multicolumn{1}{c|}{$+$} &\multicolumn{1}{c}{$-$} \\
\hline
1	&	 2&1,12	&	 2&1,12	&	 2&1,12	&	 2&1,12	\\
2	&	 7&1,3	&	 7&1,3	&	 7&1,3	&	 7&1,3	\\
3	&	 1&4,5	&	 1&4,5	&	 1&4,5	&	 1&4,5	\\
4	&	 1&21	&	 1&21	&	 1&21	&	 1&21	\\
5	&	 1&6,21	&	 1&6,21	&	 1&6,21	&	 1&6,21	\\
6	&	 1&6,21	&	 1&6,21	&	 1&6,21	&	 1&6,21	\\
7	&	 8,9&1	&	 9&1,8	&	 8&1,9	&	 9&1,8	\\
8	&	 1&21	&	 1&21	&	 1&21	&	 1&21	\\
9	&	 11&1,10,21	&	 11&1,10,21	&	 1&10,11,21	&	 11&1,10,21	\\
10	&	 11&1,10,21	&	 11&1,10,21	&	 1&10,11,21	&	 11&1,10,21	\\
11	&	 1&10,11,21	&	 21&1,10,11	&	 1&10,11,21	&	 21&1,10,11	\\
12	&	 1&13,17	&	 1&13,17	&	 1&13,17	&	 1&13,17	\\
13	&	 1&14,15	&	 1&14,15	&	 1&14,15	&	 1&14,15	\\
14	&	 1&21	&	 1&21	&	 1&21	&	 1&21	\\
15	&	 1&16,21	&	 1&16,21	&	 1&16,21	&	 1&16,21	\\
16	&	 1&16,21	&	 1&16,21	&	 1&16,21	&	 1&16,21	\\
17	&	 1&18,19	&	 1&18,19	&	 1&18,19	&	 1&18,19	\\
18	&	 1&21	&	 1&21	&	 1&21	&	 1&21	\\
19	&	 1&20,21	&	 1&20,21	&	 1&20,21	&	 1&20,21	\\
20	&	 1&20,21	&	 1&20,21	&	 1&20,21	&	 1&20,21	\\
\hline
  \end{tabular}
  \caption{A complete list of $V_{kj}(\mathcal F_{kt})$ for the second task of the TICKETS unit. Each row corresponds to a current state. The information statuses A-D are determined by the event history $\mathcal F_{kt}$, which are based on whether the fare of a concession daily subway ticket and that of four concession individual subway tickets are known. Status A: $Y_{kn}\neq 8,11$, $\forall n$, $T_{kn} < t$;
  Status B: $Y_{kn}\neq 11$, $\forall n$, $T_{kn} < t$, and there exists $m$ satisfying $T_{km} < t$ and $Y_{km} = 8$;
  Status C: $Y_{kn}\neq 8$, $\forall n$, $T_{kn} < t$, and there exists $m$ satisfying $T_{km} < t$ and $Y_{km} = 11$;
  Status D: there exists $m,n$ satisfying $T_{kn}, T_{km} < t$, $Y_{kn} = 8$, and $Y_{km} = 11$.
  The rows indicated by "$+$" show the effective actions given the corresponding current state and the information status and the rows indicated by "$-$" show the ineffective actions. }\label{tab:tickets3}

\end{table}

\begin{table}[H]
\footnotesize
  \centering
  \begin{tabular}{cccccc|ccc}
    \hline
State ID &Network&Fare&Ticket&Number&End & $+$ &$-$\\
\hline
1&NULL&NULL&NULL&NULL&0&11&1,2\\
2&CITY SUBWAY&NULL&NULL&NULL&0&1&3,7\\
3&CITY SUBWAY&FULL&NULL&NULL&0&1&4,5\\
4&CITY SUBWAY&FULL&DAILY&NULL&0&1&21\\
5&CITY SUBWAY&FULL&INDIVIDUAL&NULL&0&1&6,21\\
6&CITY SUBWAY&FULL&INDIVIDUAL&1/2/3/4/5&0&1&6,21\\
7&CITY SUBWAY&CONCESSION&NULL&NULL&0&1&8,9\\
8&CITY SUBWAY&CONCESSION&DAILY&NULL&0&1&21\\
9&CITY SUBWAY&CONCESSION&INDIVIDUAL&NULL&0&1&10,21\\
10&CITY SUBWAY&CONCESSION&INDIVIDUAL&1/2/3/4/5&0&1&10,21\\
11&COUNTRY TRAIN&NULL&NULL&NULL&0&12&1,17\\
12&COUNTRY TRAIN&FULL&NULL&NULL&0&14&1,13\\
13&COUNTRY TRAIN&FULL&DAILY&NULL&0&1&21\\
14&COUNTRY TRAIN&FULL&INDIVIDUAL&NULL&0&16&1,15,21\\
15&COUNTRY TRAIN&FULL&INDIVIDUAL&1/3/4/5&0&16&1,15,21\\
16&COUNTRY TRAIN&FULL&INDIVIDUAL&2&0&21&1,15,16\\
17&COUNTRY TRAIN&CONCESSION&NULL&NULL&0&1&18,19\\
18&COUNTRY TRAIN&CONCESSION&DAILY&NULL&0&1&21\\
19&COUNTRY TRAIN&CONCESSION&INDIVIDUAL&NULL&0&1&20,21\\
20&COUNTRY TRAIN&CONCESSION&INDIVIDUAL&1/2/3/4/5&0&1&20,21\\
21&NULL&NULL&NULL&NULL&1&NULL& NULL\\
\hline
  \end{tabular}
  \caption{A complete list of the 21 states of the first task of the TICKETS unit, and the corresponding
  effective and ineffective action types, where the effective action types are shown in the column ``$+$" and the ineffective ones are shown in the column ``$-$".  Since the current task is relatively simple, the effectiveness of event types only depends on the
  current state. }\label{tab:tickets4}
\end{table}

\end{document}